\documentclass[article,reprint,amsmath,amssymb,superscriptaddress,longbibliography]{revtex4-2}%
\usepackage{graphicx}
\usepackage{dcolumn}
\usepackage{bm}

\usepackage[colorlinks,
            linkcolor=blue,
            anchorcolor=blue,
            citecolor=blue,
            urlcolor=blue,
            ]{hyperref}
\usepackage{breakurl}
\usepackage{multirow}
\usepackage{enumitem}
\usepackage{tablefootnote}
\usepackage{threeparttable}
\usepackage{tabularx}
\usepackage{ulem}
 \setcitestyle{super}

\usepackage{booktabs}

\usepackage{lineno}

\begin{document}

\title{Quantum Geometric Origin of Strain-Tunable Giant Second-Harmonic Generation in Bi$_2$O$_2$X (X=S, Se, Te)}
	

\author{Zhefeng Lou$^{1,2,3,\dag}$, Zhihao Gong$^{4\dag}$, Ziye Zhu$^{5,6}$, Wenbin Li$^{7}$, Xiao Lin$^{2,3,*}$, and Hua Wang$^{1*}$  
~\\
	{\small\itshape{$^1$Center for Quantum Matter, School of Physics, Zhejiang University, Hangzhou 310058, China. 
	\\ 
    $^2$Key Laboratory for Quantum Materials of Zhejiang Province, Department of Physics, School of Science and Research Center for Industries of the Future, Westlake University, Hangzhou 310030, P. R. China.
    \\
    $^3$Institute of Natural Sciences, Westlake Institute for Advanced Study, Hangzhou 310024, P. R. China.
    \\
    $^4$Academy of Interdisciplinary Studies on Intelligent Molecules, Tianjin Key Laboratory of Structure and Performance for Functional Molecules, College of Chemistry, Tianjin Normal University, Tianjin, 300387, P. R. China.
        \\  
    $^{5}$Eastern Institute for Advanced Study, Eastern Institute of Technology, Ningbo, Zhejiang 315200, P. R. China.\\
    $^6$International Center for Quantum Design of Functional Materials (ICQD), and Hefei National Laboratory, University of Science and Technology of China, Hefei, 230026, P. R. China.\\
    $^7$Key Laboratory of 3D Micro/nano Fabrication and Characterization of Zhejiang Province, School of Engineering, Westlake University, Hangzhou 310024, Zhejiang Province, P. R. China.
    \\
    These authors contributed equally$^\dag$: Zhefeng Lou, Zhihao Gong. \\$*$e-mail:	linxiao@westlake.edu.cn (X.L.); daodaohw@zju.edu.cn (H.W.)}}
    }
 
	\date{\today}
	\begin{abstract}
	\noindent  Two-dimensional (2D) materials with giant nonlinear optical (NLO) responses are essential for the development of advanced on-chip NLO devices. Using first-principles calculations, we predict a remarkable strain-induced enhancement of second-harmonic generation (SHG) in the high-performance 2D semiconductors Bi$_2$O$_2$X (X = S, Se, Te). The SHG susceptibilities of Bi$_2$O$_2$X under strain are on the order of 1~nm/V, rivalling the highest values reported among 2D materials. This giant SHG response originates from gauge-invariant geometric quantities, including the quantum metric,  shift vector, and triple phase product. The strain also induces a bandgap variation in Bi$_2$O$_2$X. Intriguingly, in Bi$_2$O$_2$Te, strain-induced bandgap tuning drives a transition from a semiconductor to a half-metal, and ultimately to a polar metal. Our findings present a unique platform that combines strain-tunable bandgap engineering with exceptional NLO properties, while also highlighting the crucial role of quantum geometry in enhancing SHG.
    \end{abstract}
\maketitle

\section{Introduction}


Second-harmonic generation (SHG), a nonlinear optical (NLO) process involving frequency doubling of light, plays a crucial role in various NLO devices, including optical parametric amplifiers (OPAs)~\cite{Cerullo2021NaturePhotonics,Radic2012IEEEJ.Sel.Top.QuantumElectron.}, optical modulators~\cite{Wang2016NaturePhotonics,Qin2018ACSPhotonics}, and quantum light sources~\cite{Andrew_T.S2023Nature,Morandotti2019NaturePhotonics}. Two-dimensional (2D) NLO materials with giant SHG, such as NbOX$_2$ (X= I, Cl)~\cite{GiantNbOI2,Andrew_T.S2023Nature}, rhombohedral boron nitride (rBN)~\cite{Liu_Kaihui2024AM}, group IV monochalcogenides~\cite{wang2017giant} and transition metal dichalcogenides (TMDCs)~\cite{Zhao_Hui2013PRBMoS2500,Liu2024Science}, host promising for various applications of on-chip nonlinear optics. Notably, SHG is inherently rooted in broken inversion symmetry. Furthermore,  non-centrosymmetric phase transitions induced by external stimuli enable the design of tunable NLO switches~\cite{Liu2024NanoLett,ElectrialcontrolWSe2,ZhangX2017Nature,Zhang2021NatureElectronics}.


Bi$_2$O$_2$Se, a high-performance semiconductor with exceptional properties~\cite{ChenYulin2018Sci.adv.,Peng2017Nat.Nanotechnol.,Peng2023Nat.Mater.,tan2DFinFieldeffect2023,Peng2022Nat.Electron.,Peng2020Nat.Electron.}, undergoes a strain-induced ferroelectric (FE) phase transition~\cite{WuMH2017,ZhuZiye2024J.Mater.Chem.C,LiWB2022JACS}. Our recent work demonstrated that this transition results in a giant SHG response, an order of magnitude higher than that of NbOX$_2$~\cite{LinAdvancedMaterials2024}. These findings spark interest in exploring the whole 2D Bi$_2$O$_2$X family, featured by a lattice structure with alternating [Bi$_2$O$_2$] and [X] layers, light electron effective mass, and remarkable environmental stability~\cite{Zhai2021InfoMat,Hu2022ACSAppl.Mater.Interfaces}.


In this work, we investigate the strain-induced structural phase transition in  Bi$_2$O$_2$X using first-principles calculations, determining critical strains ($\varepsilon_\text{c}$) of approximately $3.0\%$ for X = S, Se, and Te. The indirect bandgap progressively decreases with increasing strain ($\varepsilon$) up to $\varepsilon_\text{c}$, but shows a subsequent increase beyond $\varepsilon_\text{c}$, highlighting the significant impact of this transition on the electronic structure. Intriguingly, in Bi$_2$O$_2$Te, the bandgap evolution leads to a sequential transition from semiconductor to half-metal, and finally to a polar metal.  

Furthermore, the strain-driven transition gives rise to a significant SHG response in Bi$_2$O$_2$X, with an SHG susceptibility ($\chi^{(2)}$) on the order of 1 nm/V, on par with the highest values reported in 2D materials~\cite{GiantNbOI2,Andrew_T.S2023Nature,Zhao_Hui2013PRBMoS2500,Liu2017Adv.Mater.3R,Liu2024Science}. To elucidate the microscopic and geometric mechanism, we examine the contribution of key geometric quantities to SHG, including the shift vector~\cite{wang2019ferroicity,wang2022generalized,bhalla2022PRL,BrabecPRL2023}, the quantum metric~\cite{wang2022generalized,bhalla2022PRL}, and the triple phase product of inter-band Berry connections~\cite{BrabecPRL2023}. A detailed $\boldsymbol{k}$-space analysis reveals the intricate relationship between the band structure, quantum geometry, and the SHG. These findings not only predict an extraordinary SHG response with a quantum geometric origin in Bi$_2$O$_2$X but also suggest a promising route towards developing 2D NLO semiconductors with tunable bandgaps.

\begin{figure*}[!thb]
\includegraphics[width=18cm]{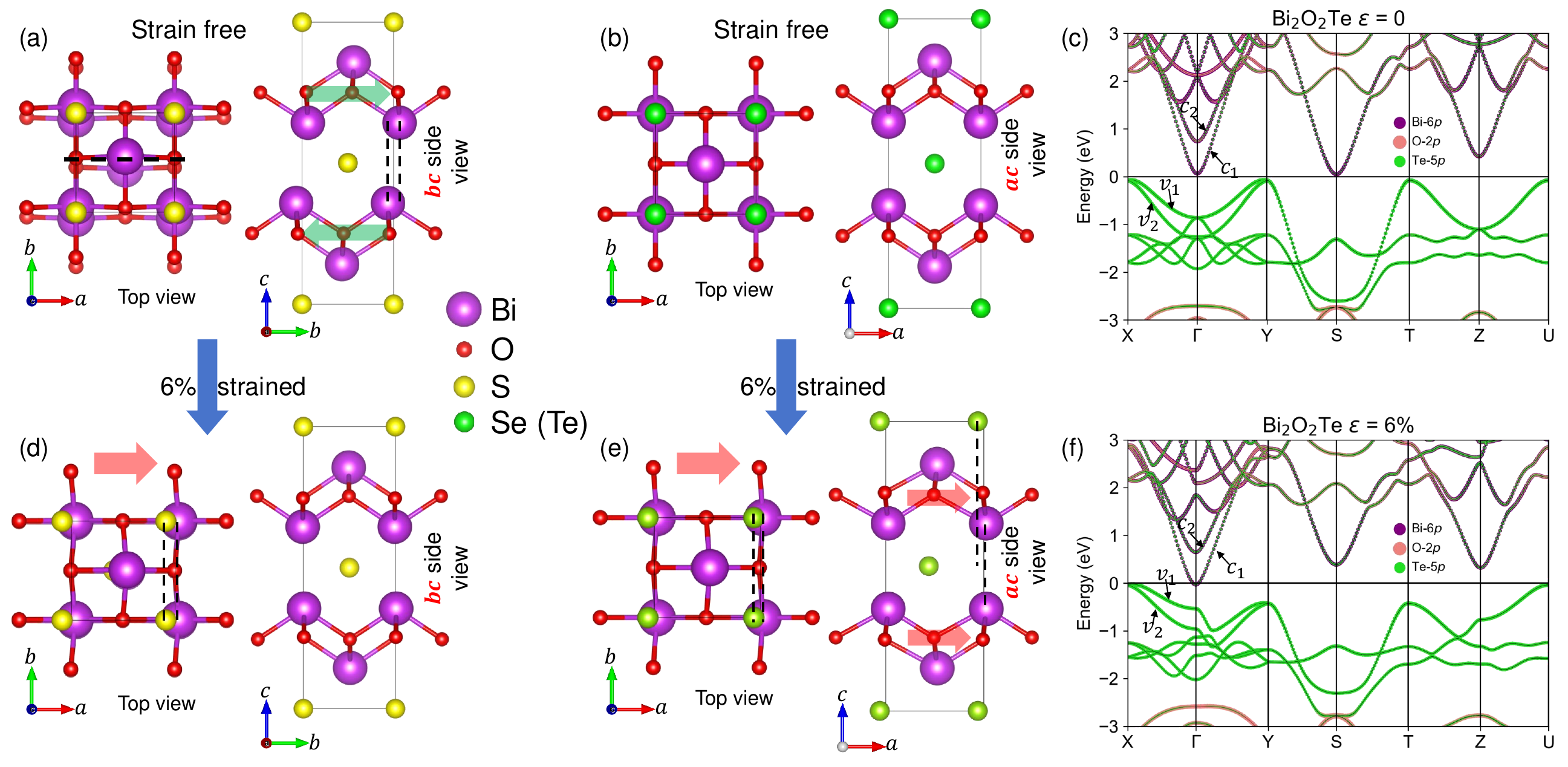}
		\renewcommand{\figurename}{FIG.}
\caption{Crystal and band structures of Bi$_2$O$_2$X under unstrained and 6\% strained conditions.
(a, b) Crystal structures of strain-free Bi$_2$O$_2$S and Bi$_2$O$_2$Se (Te), respectively. Green arrows indicate the directions of the spontaneous polarization induced by the lattice distortion of the [Bi$_2$O$_2$] frame. Dashed lines highlight these lattice distortions. The side views in (a) and (b) display the crystal structure in the $ac$-plane and $bc$-plane, respectively. (c) Band structure of strain-free Bi$_2$O$_2$Te, with orbital contributions from Bi-6$p$, O-2$p$, and Te-5$p$ orbitals. The size of circles represents the relative contribution of each orbital to the bands. The band structures of Bi$_2$O$_2$S and Bi$_2$O$_2$Se are provided in Fig.~S2. (d, e) Crystal structures of Bi$_2$O$_2$S and Bi$_2$O$_2$Se (Te) under 6\% strain, respectively. In (d, e), the red arrows indicate the net polarization arising from the shift between the [Bi$_2$O$_2$] and [Se] layers along the a-axis (marked by dashed lines). (c) Band structure of Bi$_2$O$_2$Te under 6\% strain.
}
\label{Fig1}
\end{figure*}

\section{Computational Methods}

 The crystal structures of Bi$_2$O$_2$X under various strains were calculated using density functional theory (DFT)~\cite{hohenberg1964PR, kohn_1965PR} via the Vienna {\it Ab initio} Simulation Package (VASP)~\cite{kresse1993PRB, kresse1996CMS}. The exchange-correlation energy of valence electrons was treated within the generalized gradient approximation (GGA)~\cite{becke1988PRA, langreth1983PRB} using the Perdew-Burke-Ernzerhof (PBE) parameterization~\cite{perdew1996PRL}. A plane-wave energy cutoff of 600 eV was used to ensure high accuracy in the representation of the electronic wavefunctions. A $13 \times 13 \times 4$ Monkhorst-Pack $\boldsymbol{k}$-point grid was employed for Brillouin zone sampling. Structural relaxation was deemed converged when residual forces on the ions were less than $10^{-3}$~eV/\AA. To enhance agreement with experimental electronic bandgaps, the modified Becke-Johnson (MBJ) exchange potential~\cite{PRLMBJ}, with CMBJ = 1.07, was used, as justified in Fig.~S1. SHG susceptibility calculations were done by employing a Wannier tight-binding model~\cite{marzari2012RMP, ibanez-azpiroz2018PRB, garcia2023PRB}, constructed using the Wannier90 code~\cite{pizzi2020JPCM}. A dense $90 \times 90 \times 30$ $\boldsymbol{k}$-point mesh was used for for Brillouin zone integration, with a smearing width of $\eta = 0.06$ eV applied to the Dirac $\delta$-functions. The decomposed and $\boldsymbol{k}$-resolved SHG contributions, along with quantum geometric quantities, were calculated using a modified version of the Wannier90 code with the same smearing width as for the SHG susceptibility calculations. A 3D $90 \times 90 \times 30$ $\boldsymbol{k}$-point mesh was used for decomposed SHG contributions. For $\boldsymbol{k}$-resolved quantities within the $k_z = 0$ plane, a $200 \times 200$ 2D $\boldsymbol{k}$-point mesh was employed.

\begin{figure*}[!ht]
	\includegraphics[width=18cm]{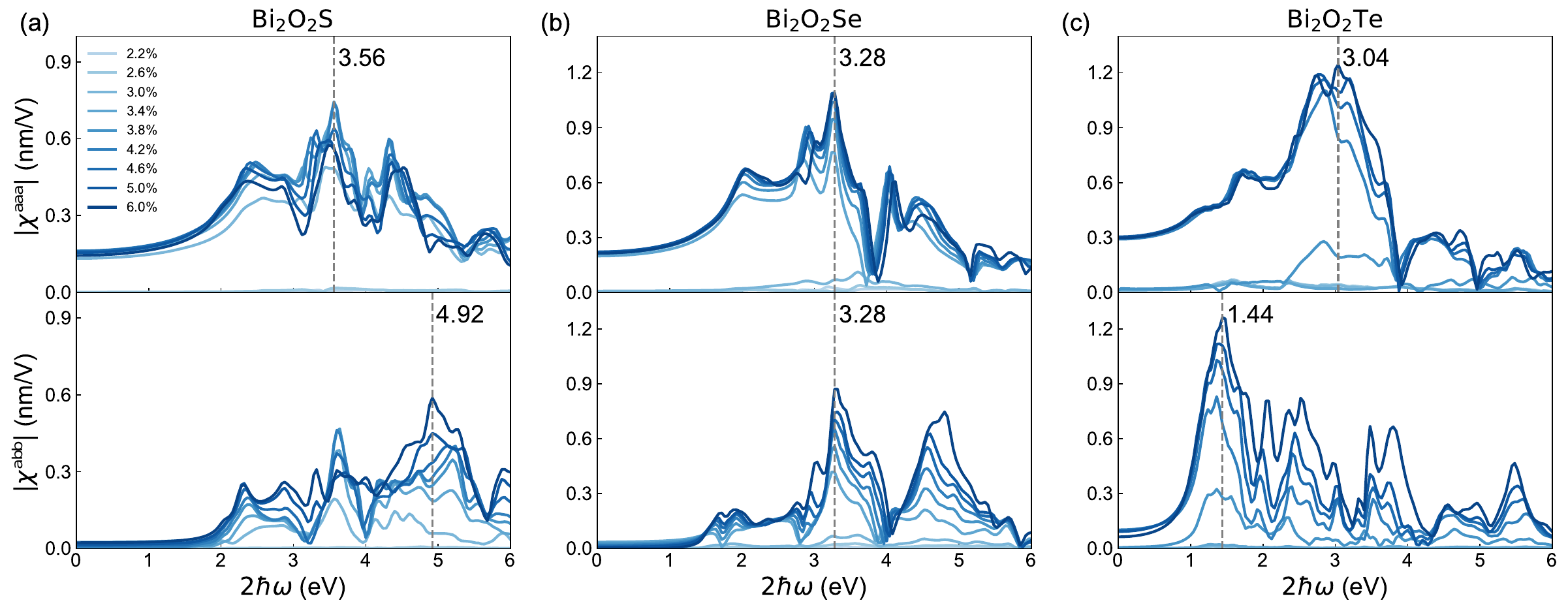}
	\vspace{-7 mm}
	\renewcommand{\figurename}{FIG.}
	\caption{Evolution of the major components of SHG susceptibility tensors in strained Bi$_2$O$_2$X. (a-c) Calculated SHG susceptibility tensors for strained Bi$_2$O$_2$S, Bi$_2$O$_2$Se, and Bi$_2$O$_2$Te, respectively. The first row in each panel represents $|\chi^{aaa}|$, while the second row displays $|\chi^{abb}|$. In all panels, the color gradient indicates the magnitude of the applied strain, with darker blue indicating higher strain levels. Gray vertical lines mark the strain values at which the maximum SHG susceptibility is achieved.}
	\label{Fig2}
\end{figure*}

\section{Results}
\subsection{Strain Tunable Structural and Electronic Properties of Bi$_2$O$_2$X}
 As illustrated in Fig.~\ref{Fig1}(a), (b), Bi$_2$O$_2$X compounds are layered materials consisting of two positively charged [Bi$_2$O$_2$] layers, sandwiched between three negatively charged [X] layers. Specifically, Bi$_2$O$_2$S crystallizes in an orthorhombic structure with $Pnnm$ space group (point group $D_{2h}$ with inversion symmetry). In Bi$_2$O$_2$S, the [Bi$_2$O$_2$] layers experience a slight distortion, resulting in an in-plane polarization along the $b$-axis, as seen in the top view of Fig.~\ref{Fig1}(a). However, this polarization is antiferroelectrically coupled between adjacent layers along the $b$-axis~\cite{WuMH2017,Zhai2021InfoMat}, side view in Fig.~\ref{Fig1}(a). In Bi$_2$O$_2$Se (Te), two [Bi$_2$O$_2$] layers remain undistorted and are inversely aligned to each other along the $c$-axis (Fig.~\ref{Fig1}(b)), forming a tetragonal anti-ThCr$_2$Si$_2$ structure with $I4/mmm$ space group (point group $D_{4h}$)~\cite{Zhai2021InfoMat}.

The band structures of the three materials exhibit strong similarities. The valence bands ($v_1$, $v_2$ ...) near the bandgap are predominantly derived from the X-x$p$ orbitals (x= 3, 4, 5 for X= S, Se, Te), while the conduction bands ($c_1$, $c_2$ ...) and deeper valence bands primarily originate from the Bi-6$p$ and O-2$p$ orbitals, respectively, as illustrated in Fig.~\ref{Fig1}(c) and Fig.~S2(b, c). In Bi$_2$O$_2$Se and Bi$_2$O$_2$Te, the presence of $C_4$ rotational symmetry along the $c$-axis ensures the identical band structures along the $\Gamma$-X and $\Gamma$-Y high-symmetry paths. In Bi$_2$O$_2$S, the antiferroelectric arrangement along the $b$-axis breaks the $C_4$ symmetry, resulting in slight differences in the band structures along these two paths, as shown in Fig.~S2(c).

\begin{table}[!ht]
\renewcommand{\arraystretch}{1.5}
\setlength{\tabcolsep}{1.6mm}
\caption{Initial crystal structures of Bi$_2$O$_2$X and their corresponding $\varepsilon_\text{c}$}
    \centering
    \begin{tabular}{llllll}
    \hline
        ~ & Space group & $|a|$ (\AA) & $|b|$ (\AA) & $|c|$ (\AA) & $\varepsilon_\text{c}$ (\%) \\ 
        \hline
        Bi$_2$O$_2$S & $Pnnm$ & 3.840 & 3.874 & 11.920 & 3.0 \\ 
        Bi$_2$O$_2$Se & $I4/mmm$ & 3.880 & 3.880 & 12.160 & 3.0 \\ 
        Bi$_2$O$_2$Te & $I4/mmm$ & 3.980 & 3.980 & 12.700 & 3.6 \\ 
        \hline
    \end{tabular}
    \label{table1}
\end{table}

Consistent with previous studies on strain-induced ferroelectricity in Bi$_2$O$_2$Se~\cite{WuMH2017,ZhuZiye2024J.Mater.Chem.C,LiWB2022JACS}, Bi$_2$O$_2$X undergoes a phase transition from a centrosymmetric to a non-centrosymmetric structure under uniaxial strain.  Table~\ref{table1} summarises the $\varepsilon_\text{c}$ for these transitions, alongside the space groups and lattice constants of the pristine Bi$_2$O$_2$X structures.
Application of a 6\% strain along the $a$-axis induces relative displacements between the [Bi$_2$O$_2$] and [X] layers along this axis in all three materials, as shown in Fig.~\ref{Fig1}(d, e).  
In the case of Bi$_2$O$_2$S, this applied strain also relieves the $b$-axis offset of the [Bi$_2$O$_2$] layers, thus weakening the inherent antiferroelectric order along this direction. 
While this behavior differs from Bi$_2$O$_2$Se and Bi$_2$O$_2$Te (when $\varepsilon>\varepsilon_\text{c}$), a net polarization emerges along the $a$-axis in all three materials, resulting in a $C_{2v}$ symmetry.

In Bi$_2$O$_2$X, Strain-induced structural distortions significantly modify the electronic band structures and associated geometric quantities.
As depicted in Figs.~\ref{Fig1}(c, f), and S2(d), the indirect bandgaps exhibit a non-monotonic dependence on strain, narrowing to a minimum at $\varepsilon_\text{c}$, 
and then widening again with increasing strain. The induced polarization breaks the inherent 4-fold rotational symmetry in Bi$_2$O$_2$Te and Bi$_2$O$_2$Se, leading distinct changes in the band structure along the $\Gamma$-X and $\Gamma$-Y high-symmetry paths. In Bi$_2$O$_2$S, however, asymmetry
between bands of $\Gamma$-X and $\Gamma$-Y are already present in pristine structures, and strain further amplifies this anisotropy,
as shown in Fig.~S2(f). It is noteworthy that our calculations predict a transition from a semiconducting state (with a narrow bandgap of $\sim 0.12$ eV) to a half-metallic state in Bi$_2$O$_2$Te upon reaching the critical strain $\varepsilon_\text{c}$. 
The strained Bi$_2$O$_2$Te is thus likely to behave as a polar metal, considering the presence of internal polarization. These strain tunable electronic structures, ferroelectric polarizations, and the associated strong inversion symmetry-breaking in the Bi$_2$O$_2$X indicate its potential for engineering strong SHG responses.


\begin{figure*}[!th] 
	\includegraphics[width=17.5cm]{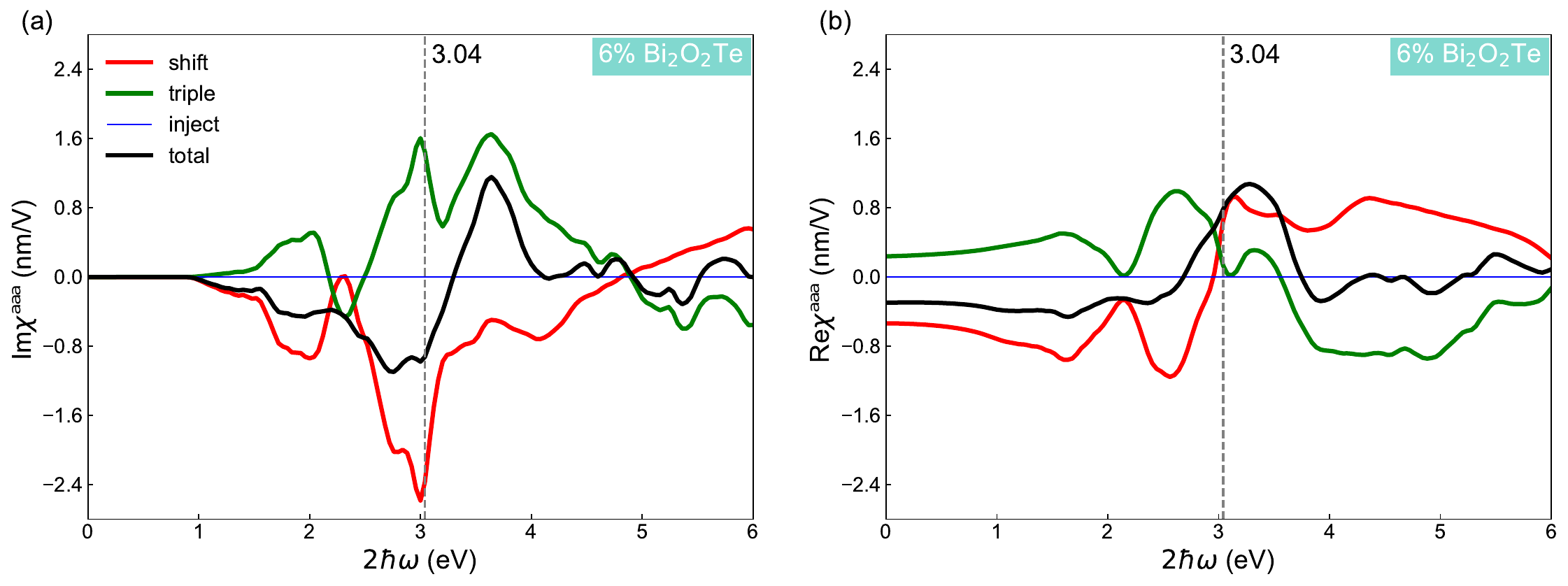}
	\vspace{-5mm}
	\renewcommand{\figurename}{Fig}
	\caption{Frequency-dependent (a) imaginary and (b) real part of the `shift' $\chi_{\text{shift}}^{aaa}$, `triple' $\chi_{\text{triple}}^{aaa}$, and `inject' $\chi_{\text{inject}}^{aaa}$ contributions to the SHG susceptibility $\chi^{aaa}$.}
	\label{Fig3}
\end{figure*}

\subsection{Strain Tunable SHG of Bi$_2$O$_2$X}

The SHG effect is described by the following equation:
\begin{equation}
P_{i}(2\omega) = \epsilon_0 \chi^{(2)}_{ijk} E_{j}(\omega) E_{k}(\omega)
\label{basicSHG}
\end{equation}
where $\boldsymbol{P}(2\omega)$ is the induced electric polarization at the second-harmonic frequency ($2\omega$),  $\epsilon_0$ is the vacuum permittivity, $\chi^{(2)}_{ijk}$ is the second-order nonlinear susceptibility tensor, and $\boldsymbol{E}(\omega)$ is the incident electric field at frequency $\omega$. The indices $i$, $j$, and $k$ represent Cartesian coordinates.

In strained Bi$_2$O$_2$X, the SHG susceptibility tensors $\chi^{(2)}_{ijk}$ contain five independent components under $C_{2v}$-symmetry, including $\chi^{(2)}_{aaa}$, $\chi^{(2)}_{abb}$, $\chi^{(2)}_{acc}$, $\chi^{(2)}_{cac} = \chi^{(2)}_{cca}$, and $\chi^{(2)}_{bab} = \chi^{(2)}_{bba}$, according to Neumann's principle and permutation symmetry. Here, the superscripts `$ijk$' are aligned with the crystal axes (`$a$', `$b$', `$c$'). 
The magnitudes of complex susceptibility components, $|\chi^{ijk}|$, are presented in Figs.~\ref{Fig2} and S3 for strains of $\varepsilon \geq 2.2$\%. 
In general, the magnitudes of SHG susceptibilities increase with strain.
The components $|\chi^{aaa}|$ and $|\chi^{abb}|$, shown in Figure~\ref{Fig2}, are dominant in SHG responses with a maximum magnitude 
reaching approximately 1~nm/V, which is 
among the highest reported for 2D materials~\cite{GiantNbOI2,Andrew_T.S2023Nature,Zhao_Hui2013PRBMoS2500,review_2D,Zhang_2020_review_2D,Liu2017Adv.Mater.3R,Liu2024Science}. Within the strain range, $2.2\%\leq\varepsilon\leq 6\% $, 
the component $|\chi^{aaa}|$ for Bi$_2$O$_2$S and Bi$_2$O$_2$Se 
decrease when strain exceeds $\varepsilon = 3.8$\% and 5\% , respectively (details in Fig.~S4),
while  
$|\chi^{abb}|$ and rest components in Fig.~S3
increase monotonically until saturation.
For Bi$_2$O$_2$Te, all five components increase monotonically. As shown in Fig.~\ref{Fig2}(c), the peaks of frequancy-dependent components $|\chi^{aaa}|$ and $|\chi^{abb}|$ are at approximately 1.25~nm/V at $2\hbar \omega=$ 3.04 eV and 1.44 eV, respectively.
Bi$_2$O$_2$Te thus maintains a high SHG response over a broader spectral range compared to the other two materials, suggesting its potential for broadband optical signal processing applications~\cite{Radic2012IEEEJ.Sel.Top.QuantumElectron.}.

\section{Discussion}

While the microscopic theory of SHG is well-established, the underlying quantum geometric contributions remain relatively unexplored. To gain deeper insights, we decompose the SHG susceptibilities into three distinct geometric terms according to the second-order shift and injection current responses and subsequently analyze the $\boldsymbol{k}$-resolved contributions of SHG.

\begin{figure*}[!htb]
 	\includegraphics[width=18cm]{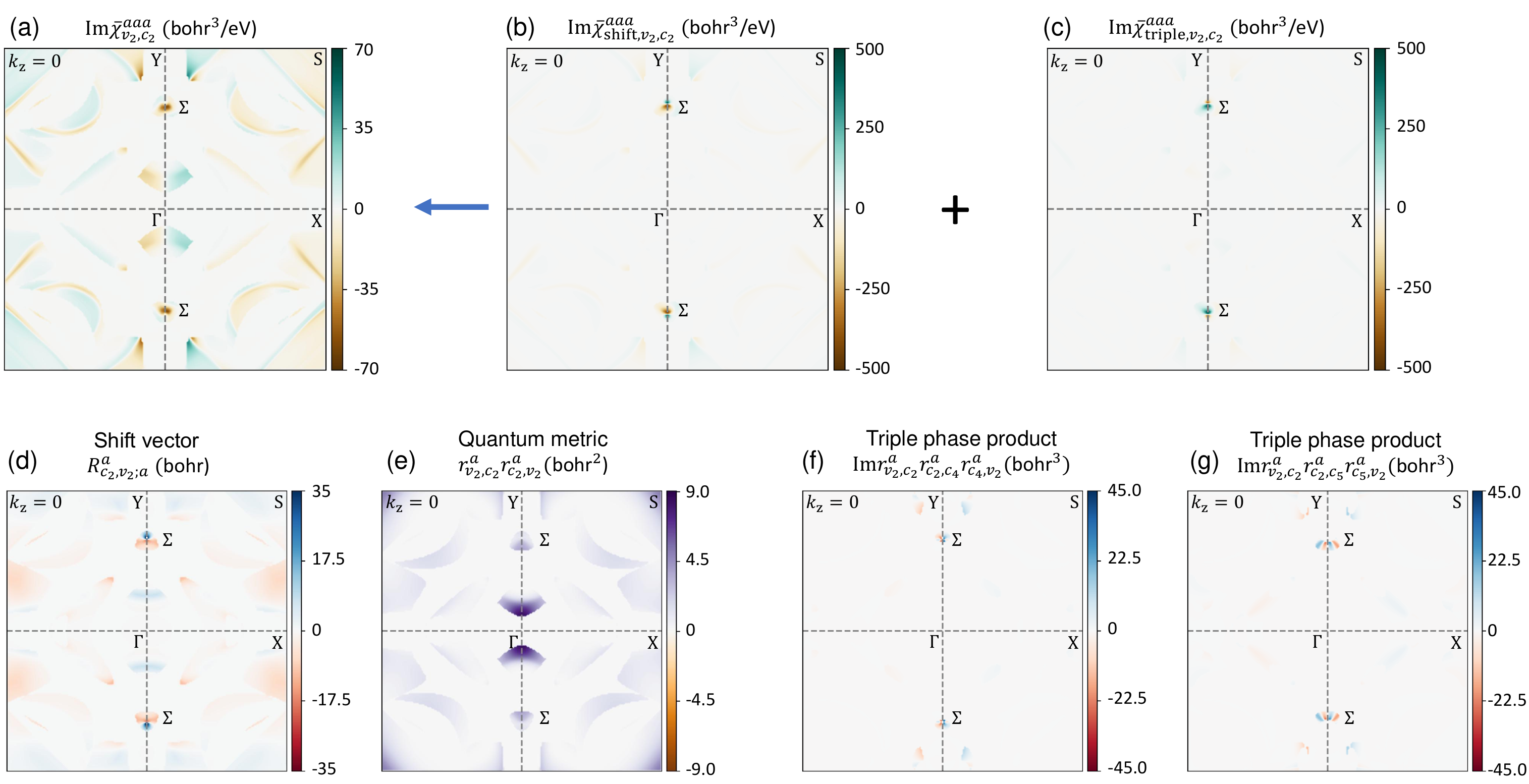}
         \vspace{-5mm}
	\renewcommand{\figurename}{FIG.}
	\caption{Microscopic origin of Im$\chi^{aaa}$ (at $\varepsilon = 6$\%) for Bi$_2$O$_2$Te. (a–c) $\boldsymbol{k}$-resolved contributions to Im$\overline{\chi}^{aaa}$, including Im$\overline{\chi}^{aaa}_{v_2,c_2}$, Im$\overline{\chi}^{aaa}_{\text{shift},v_2,c_2}$, and Im$\overline{\chi}^{aaa}_{\text{triple},v_2,c_2}$, at $k_z = 0$. Note that (a) employs a smaller scale bar than (b) and (c) to visualize the smaller magnitude of Im$\overline{\chi}^{aaa}_{v_2,c_2}$, which arises from the cancellation between Im$\overline{\chi}^{aaa}_{\text{shift},v_2,c_2}$ and Im$\overline{\chi}^{aaa}_{\text{triple},v_2,c_2}$. (d) Distribution of the shift vector, $R_{c_2,v_2;a}^{a}$. (e) $\boldsymbol{k}$-resolved quantum metric, $r_{v_2,c_2}^{a} r_{c_2,v_2}^{a}$. (f, g) $\boldsymbol{k}$-resolved triple products, $r_{v_2,c_2}^{a} r_{c_2,c_4}^{a} r_{c_4,v_2}^{a}$ and $r_{v_2,c_2}^{a} r_{c_2,c_5}^{a} r_{c_5,v_2}^{a}$, respectively. Here, $\Sigma = (0, \pm0.647, 0)$ are the points Im$\overline{\chi}^{aaa}_{\text{shift},v_2,c_2}$ and Im$\overline{\chi}^{aaa}_{\text{triple},v_2,c_2}$ concentrated.}
	\label{Fig4}
\end{figure*}
\subsection{Decomposition Scheme for SHG Susceptibilities} 

The total SHG susceptibility, $\chi^{ijk}$, can be decomposed into three contributions arising from distinct geometric quantities: the shift current term ($\chi^{ijk}_\text{shift}$), the triple phase product of the inter-band Berry connection ($\chi^{ijk}_\mathrm{triple}$), and the injection current term ($\chi^{ijk}_\mathrm{inject}$). The expressions for these contributions are given below:
\begin{equation}
	\begin{aligned}
    \chi_{\text{shift}}^{{ijk}} &=  \frac{ie^3}{2\hbar^2V} \sum_{{nm}\boldsymbol{k}}
      f_{{nm}}\Bigg[  \frac{2r_{{nm}}^{i}\left(r_{{mn};{k}}^{{j}}+r_{{mn};{j}}^{{k}}\right)}{\omega_{{mn}}(\omega_{{mn}}-2\tilde{\omega})}\\   
      &+\frac{r_{{nm};{j}}^{{i}}r_{{mn}}^{{k}}+r_{{nm};{k}}^{{i}}r_{{mn}}^{{j}} } {\omega_{{mn}}(\omega_{{mn}}-\tilde{\omega})}
      -\frac{r_{{nm};{i}}^{{k}}r_{{mn}}^{{j}}+r_{{nm};{i}}^{{j}}r_{{mn}}^{{k}}}{2\tilde{\omega}(\omega_{{mn}}-\tilde{\omega})} \Bigg]
	\end{aligned}
     \label{Eqshift}
  \end{equation}
  
\begin{equation}
	\begin{aligned}
	\chi_{\text{triple}}^{{ijk}} &=\frac{e^3}{2\hbar^2V}\sum_{{nlm}\boldsymbol{k}}f_{nm}\Bigg[\frac{2r_{nm}^{i}\left(r_{ml}^{j} r_{ln}^{k}+r_{ml}^{k}r_{ln}^{j}\right)}{(\omega_{ln}-\omega_{ml})(\omega_{mn}-2\tilde{\omega})}\\
 &+\frac{r_{ml}^{i}\left(r_{ln}^{j} r_{nm}^{k}+r_{ln}^{k}r_{nm}^{j}\right)}{(\omega_{nm}-\omega_{ln})(\omega_{nm}-\tilde{\omega})}+\frac{r_{ln}^{i}\left(r_{nm}^{j} r_{ml}^{k}+r_{nm}^{k}r_{ml}^{j}\right)}{(\omega_{ml}-\omega_{nm})(\omega_{nm}-\tilde{\omega})}\Bigg]
	\end{aligned}
	\label{Eqtriple}
\end{equation}  

\begin{equation} 
	\begin{aligned}
    \chi_{\text{inject}}^{{ijk}} &=\frac{ie^3}{2\hbar^2V}
 \sum_{nm\boldsymbol{k}} f_{nm}\Bigg[\frac{r_{nm}^{{i}}\left(r_{mn}^{{j}}\Lambda_{mn}^{{k}}+r_{mn}^{{k}}\Lambda_{mn}^{{j}}\right)}{\omega_{mn}^{2}}
  \\&\cdot\left(\frac{1}{\omega_{mn}-\tilde{\omega}}-\frac{4}{\omega_{mn}-2\tilde{\omega}}\right)-\frac{\Lambda_{nm}^{{i}}\left(r_{mn}^{{j}}r_{nm}^{{k}}+r_{mn}^{{k}}r_{nm}^{{j}}\right)}{4\tilde{\omega}^2(\omega_{mn}-\tilde{\omega})}
	\Bigg]
	\end{aligned} 
  \label{Eqinject}
\end{equation}

The 'shift' term exhibits similarities to the shift current response and is associated with the geometric shift vector and quantum metric. The 'triple' term is directly linked to the geometric triple phase product, a quantity akin to Bargmann invariants. The 'inject' term, analogous to the injection current, is associated with the quantum geometric tensor and band asymmetry. In these expressions, $\tilde{\omega} = \omega + i\eta/\hbar$ represents the fundamental frequency with a small imaginary smearing factor $\eta$ introduced for numerical calculations. Here, $\omega_{nm} \equiv \omega_{n} - \omega_{m}$ and $f_{nm} \equiv f_{n} - f_{m}$, where $\hbar\omega_{n}$ represents the eigenvalue and $f_{n} = f(\hbar\omega_{n})$ denotes the occupation factor of the Bloch eigenstate $|n,\boldsymbol{k}\rangle$. Note that $\boldsymbol{k}$ represents a vector in reciprocal space, distinct from the Cartesian index $k$. The inter-band Berry connection for bands $n \ne m$ is given by $r_{nm}^{i} = i\langle n,\boldsymbol{k} | \partial_{i} | m,\boldsymbol{k}\rangle$.  Its covariant derivative is defined as $r_{nm;j}^{i} = \partial_{j} r_{nm}^{i} - i(A_{n}^{j} - A_{m}^{j})r_{nm}^{i}$, where $A_{n}^{j} = i\langle n,\boldsymbol{k} | \partial_{j} | n,\boldsymbol{k}\rangle$ is the intra-band Berry connection.  Finally, $\Lambda_{nm}^{i} = v_{n}^{i} - v_{m}^{i}$ represents the difference in group velocities between bands $n$ and $m$.

\subsection{Geometric Origin of SHG} 

Taking $|\chi^{aaa}|$ as an example, band-resolved and $\boldsymbol{k}$-resolved analyses reveal that both `shift' and `triple' contribute significantly to the SHG response in strained Bi$_2$O$_2$Te. 
The spectra of `shift', `triple', and `inject' contributions to SHG susceptibilities are presented in Fig.~\ref{Fig3}, where dashed lines label the maximum response for $|\chi^{aaa}|$ at 3.04~eV. Notably, the `inject' term vanishes in time-reversal symmetric systems for $|\chi^{aaa}|$, as the quantum metric $r_{nm}^{a}r_{mn}^{a}$ is time-reversal symmetric, while the group velocity difference $\Lambda_{nm}^{i}$ is time-reversal antisymmetric.

The analysis of contributions from ($v_2$, $c_2$) band pairs is representative in characterizing basic behaviors of `shift' and `triple' parts of Im$\chi^{aaa}$, as reflected in the spectrum profiles of Im$\overline{\chi}^{aaa}_{\text{shift},v_2,c_2}$  and Im$\overline{\chi}^{aaa}_{\text{triple},v_2,{c}_2}$ shown in Fig.~S5. Here, the overline, `-', represents the average over permuted band indices, 
Im$\overline{\chi}^{{aaa}}_{{v}_2,{c}_2}$ = 1/2(Im$\chi^{{aaa}}_{{v}_2,{c}_2}$ + Im$\chi^{{aaa}}_{{c}_2,{v}_2}$).
Figures~\ref{Fig4}(a)-(c) showcase distributions of 
Im$\overline{\chi}^{{aaa}}_{{v}_2,{c}_2}$, and its
 `shift' and `triple' terms are displayed over the $(k_x, k_y)$ plane at $k_z = 0$, where they are found to be mainly concentrated at $\Sigma$ points.
The opposite sign of `shift' and `triple' terms at the $\Sigma$ points coincides with spectrum of Im$\chi^{aaa}$ in Fig.~\ref{Fig3}(a).
Figure~\ref{Fig4}(a) illustrates `shift' and `triple' contributions in the $k_z = 0$ plane, which are more clearly evident in Fig.~S6. 

To gain a clear view of geometric origins in `shift' term of `$aaa$' component,
we re-formulate the $\boldsymbol{k}$-resolved contribution of $\chi_{\mathrm{shift}}^{aaa}$
into a summation of weighted fractions $D_{\text{shift}}^{{aaa}}(\tilde{\omega})$, characterizing energy resonances of $\omega_{nm}$ (see Note S1 for details), as  
\begin{equation} 
	\begin{aligned}
 \chi_{\mathrm{shift}}^{aaa}(\boldsymbol{k})&=\frac{ie^3}{\hbar^2V}\sum_{nm}f_{nm}r_{nm}^{a}r_{mn;a}^{a}D_\text{shift}^{aaa}(\tilde{\omega})
	\end{aligned} 
\label{shiftaaa1}
\end{equation}
where the weights $r_{{nm}}^{{a}}r_{{mn;a}}^{{a}}$ are decomposed into $\Gamma^{{aaa}}_{{mn}} - i\tilde{\Gamma}^{{aaa}}_{{mn}}$~\cite{bhalla2022PRL}
with the metric connection,  $\Gamma^{{aaa}}_{{mn}}$, and the symplectic connection, $\tilde{\Gamma}^{{aaa}}_{{mn}}$. 
Here, $\tilde{\Gamma}^{{aaa}}_{{mn}}$ is derived from the product of 
the shift vector ($R_{{mn;a}}^{{a}}$) and the quantum metric ($r_{{nm}}^{{a}}r_{{mn}}^{{a}}$)~\cite{AhnNatphys2022,bhalla2022PRL}.
Figures~\ref{Fig4}(d) and (e) present the $\boldsymbol{k}$-resolved distributions of $R_{{c}_2,{v}_2;{a}}^{{a}}$ and  $r_{{v}_2,{c}_2}^{{a}}r_{{c}_2,{v}_2}^{{a}}$, respectively.
Both $R_{{c}_2,{v}_2;{a}}^{{a}}$ and $r_{{v}_2,{c}_2}^{{a}}r_{{c}_2,{v}_2}^{{a}}$ exhibit distributions 
similar to that of Im$\overline{\chi}^{{aaa}}_{{v}_2,{c}_2}$, whereas the shift vector $R_{{c}_2,{v}_2;{a}}^{{a}}$ possess peak concentrations 
at the $\Sigma$ points, which in turn leads to the characteristic peaks in Im$\overline{\chi}^{{aaa}}_{{v}_2,{c}_2}$.

The `triple' term is similarly re-expressed as:
\begin{equation} 
	\begin{aligned}
		\chi_{\mathrm{triple}}^{\mathrm{aaa}}(\boldsymbol{k})&=\frac{e^3}{\hbar^2V}\sum_{{nlm}}f_{nm}r_{nm}^{a}r_{ml}^{a} r_{ln}^{a}D_\text{triple}^{aaa}(\tilde{\omega})
	\end{aligned} 
	\label{tripleaaa1}
\end{equation}
where the form of $D_{\text{triple}}^{{aaa}}(\tilde{\omega})$ can be found in 
Note S1, and the weights of the combined band indices, ($n$, $m$, $l$),   
are determined by the triple phase product of inter-band Berry connections, 
$r_{nm}^{a}r_{ml}^{a} r_{ln}^{a}$,
which is gauge invariant. The distribution of Im$\overline{\chi}^{{aaa}}_{\text{triple},v_2,c_2}$
in Fig.~\ref{Fig4}(c) is primarily sourced from contributions of 
($v_2$, $c_2$, $c_4$) and ($v_2$, $c_2$, $c_5$), both of which exhibit distributions similar to Im$\overline{\chi}^{{aaa}}_{\text{triple},v_2,c_2}$,
as shown in Figs~\ref{Fig4}(f) and (g), respectively.

For the `$abb$' component, a similar analysis is performed, revealing that 
the `shift' and `inject' contributions of Im$\chi^{abb}$ stem from geometric quantities such as $r_{{nm}}^{{a}}r_{{mn;b}}^{{b}}$, $r_{{ml}}^{{a}}r_{{ln}}^{{b}}r_{{nm}}^{{b}}$, and $r_{{nm}}^{{a}}r_{{mn}}^{{b}}\Lambda_{{mn}}^{{b}}$. Here, the `inject' term is allowed by the time-reversal antisymmetric Berry curvature $r_{{nm}}^{{a}}r_{{mn}}^{{b}}$ with $a\neq b$.
Further details can be found in Note~S1 and Fig.~S7. 
Additionally, the real parts of `$aaa$' and `$abb$' are also contributed from both
`shift' and `triple' terms at peak frequencies, shown in Figs.~\ref{Fig3}(b) and S7(b). 
Our analysis demonstrates that 
shift vector $R_{{mn;a}}^{{a}}$, quantum metric $r_{{nm}}^{{a}}r_{{mn}}^{{a}}$, and triple phase product $r_{{nm}}^{{a}}r_{{ml}}^{{a}} r_{{ln}}^{{a}}$
serve as the primary geometric origins for the $\chi^{aaa}$ component of the SHG susceptibility. A similar analysis can be applied to other components, such as $\chi^{abb}$.

\section{Conclusion}


In summary, this study reveals the exceptional tunability of SHG in Bi$_2$O$_2$X, driven by strain-induced structural phase transitions. We investigate the strong correlation between the SHG response and the quantum geometric properties of electronic wavefunctions, specifically identifying the significant contributions of the shift vector, quantum metric, and triple phase product of inter-band Berry connections to the observed strong SHG in Bi$_2$O$_2$X. Our work provides a quantum geometric perspective on the strain tunable NLO properties in 2D materials. Given its excellent environmental stability and superior electronic and optoelectronic characteristics~\cite{li2DBi2O2SeEmerging2021,tan2DFinFieldeffect2023,tongSensitiveUltrabroadbandPhototransistor2019,fuUltrasensitive2DBi2O2Se2019,Zhai2021InfoMat,Hu2022ACSAppl.Mater.Interfaces}, Bi$_2$O$_2$X holds promise as a versatile platform for advanced NLO applications, including next-generation highly sensitive and switchable NLO devices.

\section{Acknowledgments}
This research is supported by ``Pioneer'' and ``Leading Goose'' R$\&$D Program of Zhejiang under Grant 2024SDXHDX0007 and Zhejiang Provincial Natural Science Foundation of China for Distinguished Young Scholars under Grant No. LR23A040001. X.L. also acknowledges the support by National Key R$\&$D Program of China under Grant No.2024YFA1408100 and National Natural Science Foundation of China (NSFC) under Grant No. 12474131.
H.W. acknowledges the support from the NSFC under Grant Nos. 12304049 and 12474240.
W.L. acknowledges the support by Research Center for Industries of the Future at Westlake University under Award No. WU2022C041 and the NSFC under Grant No. 62374136.

%










\normalem
\bibliography{BOXSHG}



\end{document}